\pdfoutput=1

\documentclass[aps,prl,reprint,superscriptaddress,longbibliography]{revtex4-2}

\usepackage[caption=false]{subfig}
\usepackage{graphicx}
\usepackage{amsmath}
\usepackage[normalem]{ulem}
\usepackage{siunitx}
\usepackage{svg}
\usepackage{amssymb}
\usepackage{mathtools}
\usepackage[colorlinks=true,linkcolor=blue,urlcolor=blue,citecolor=blue]{hyperref}

\newcommand*\diff{\mathop{}\!\mathrm{d}}

\newcommand{\figref}[1]{Fig.~\ref{#1}}

\newcommand{\Eqref}[1]{Eq.~\eqref{#1}}
\newcommand{\Eqsref}[1]{Eqs.~\eqref{#1}}
\newcommand{\mean}[1]{\langle #1 \rangle}

\newcommand{\la}[1]{l_{\mathrm{a}, #1}}
\newcommand{\lp}[1]{l_{\mathrm{p}, #1}}
\newcommand{\kBT}{k_\mathrm{B}T}

\usepackage[caption=false]{subfig}

\newcommand{\mysection}[1]{}
\newcommand{\mysubsection}[1]{}

\begin{document}

\title{Fluctuating chemically active droplets}

\author{Guido L. A. Kusters}
\affiliation{Max Planck Institute for Dynamics and Self-Organization, Göttingen, Germany}

\author{Noah Ziethen}
\affiliation{DAMTP, Centre for Mathematical Sciences, University of Cambridge, Cambridge CB3 0WA, United Kingdom}

\author{Michael E. Cates}
\affiliation{DAMTP, Centre for Mathematical Sciences, University of Cambridge, Cambridge CB3 0WA, United Kingdom}

\author{David Zwicker}
\affiliation{Max Planck Institute for Dynamics and Self-Organization, Göttingen, Germany}

\begin{abstract}
    We study size fluctuations of chemically active droplets in the regime of weak reactions and strong phase separation.
    By combining a thin-interface model for the deterministic droplet growth with stochastic noise based on fluctuation-dissipation arguments, we report excellent quantitative agreement between our theory and numerical field simulations.
    Finally, we use our model to infer the effective non-equilibrium potential that governs droplet growth, and show that active reactions that enable size control suppress nucleation.
\end{abstract}

\maketitle

\mysection{Introduction}\label{sec:introduction}

Chemically active droplets combine phase separation and non-equilibrium reactions, with particular relevance to biology \cite{hyman2014liquid,banani2017biomolecular,alberti2017wisdom,berry2018physical,lasker2022material,klosin2020phase,deviri2021physical,zechner2025concentration,Monchaux2025buffering,brangwynne2009germline,wheeler2016distinct,yoo2022chaperones} and engineered synthetic systems \cite{ge2009vivo,dzuricky2020novo,tomares2022repurposing,bergmann2023liquid,liao2024charge,koga2011peptide,saha2019condensates,kuffner2020acceleration,peeples2021mechanistic,smokers2024droplets,deng2020programmable,donau2023chemistry,lv2015photocatalytic,jambon2024phase}.
Noise is relevant in these systems because droplets are typically small and soft \cite{terentjev2015oxford}, which results in significant size fluctuations, but also for the study of nucleation.
While both of these aspects have been studied extensively in passive droplets \cite{safran1983fluctuations,milner1987dynamical,farago1990shape,buff1965interfacial,weeks1977structure,aarts2004direct,langer1968theory,langer1969statistical}, and in special cases for active droplets \cite{mabillard2023heat,kumar2023fluctuations,bauermann2025theory,cho2023tuning,cho2023nonequilibrium,ziethen2023nucleation,cho2025interfacial,fausti2024statistical,prathyusha2024anomalous,fausti2021capillary,cates2023classical,tjhung2018cluster}, there currently exists no comprehensive theory for fluctuating chemically active droplets.

In this Letter, we study fluctuating chemically active droplets in the regime of weak reactions and strong phase separation.
In this case, the interface dynamics is governed by local diffusive fluxes, as in Model~B~\cite{hohenberg1977theory}.
However, in contrast to Model~B, reactions modify those fluxes and break material conservation.
In particular, sustained energy input in chemically active droplets can lead to non-equilibrium stationary states with finite droplet sizes.
We address fluctuations in this regime by deriving an effective Langevin equation for the droplet radius that combines the full deterministic dynamics with noise derived in the passive limit.
Our theory predicts droplet size fluctuations in excellent quantitative agreement with numerical field simulations, and describes when active reactions suppress nucleation.

\mysection{Size of chemically active droplets fluctuates}\label{sec:system}

\begin{figure}
\centering
\includegraphics[width=\columnwidth]{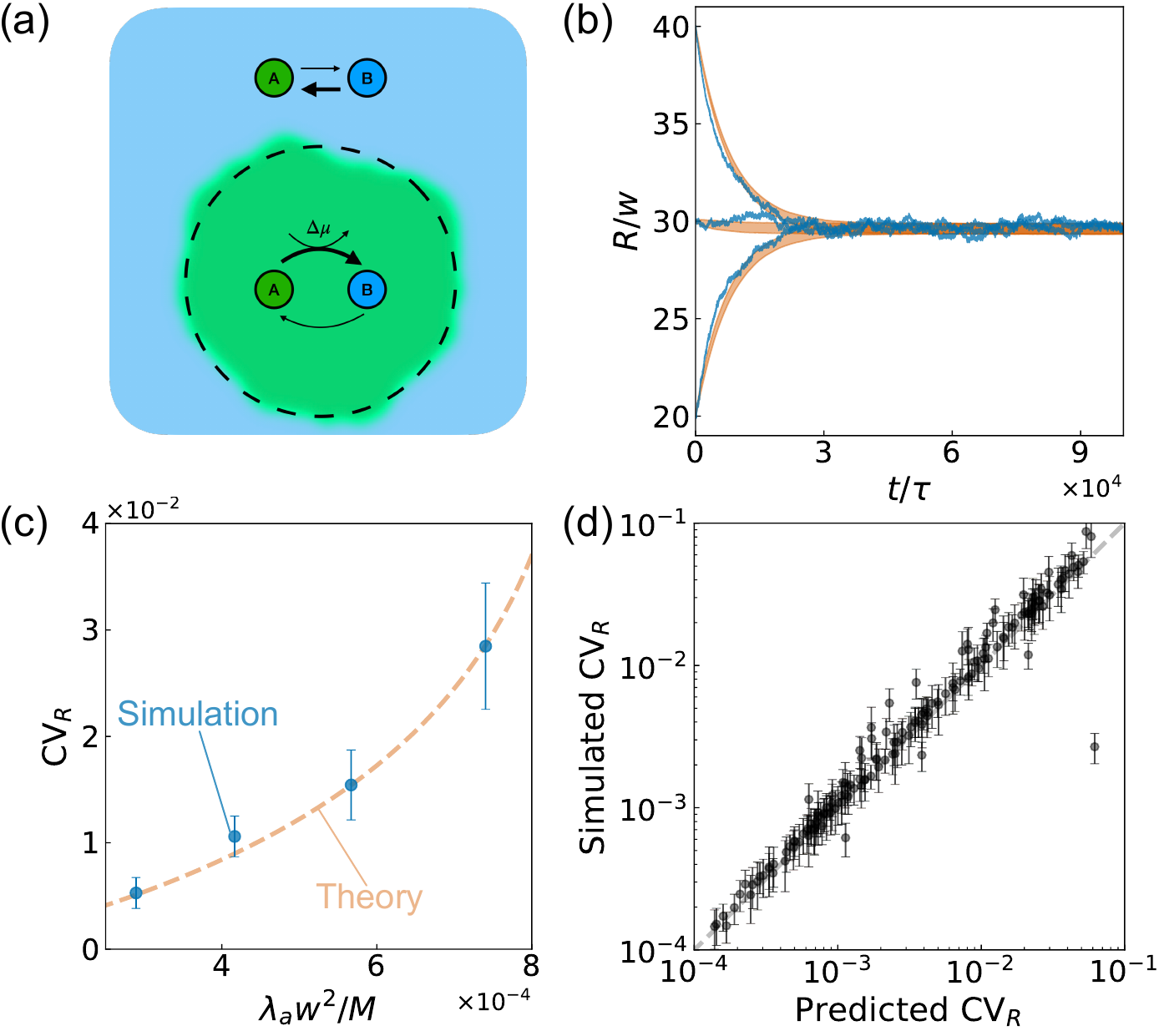}
\caption{\textbf{Chemically active droplets with fluctuations.}
(a) Schematic of components A and B phase separating and converting into each other with external energy input~$\Delta\mu$.
(b) Droplet radius~$R$ as a function of time~$t$ from numerical simulations (blue lines) and our theoretical prediction (orange shade, one standard deviation confidence interval) for three initial sizes, where $\tau=\kappa/Mb^2$.
(c) Steady-state coefficient of variation $\mathrm{CV}_R=\sqrt{\mean{R^2}-\mean{R}^2}/\mean{R}$ as a function of the active reaction rate~$\lambda_\mathrm{a}$ from numerical simulations (symbols) and theoretical prediction (dashed line).
(d) Simulated $\mathrm{CV}_R$ compared to theoretical predictions (black disks) exhibit strong correlations (dashed line).
\mbox{(b--d)}
Theoretical prediction is $\mathrm{CV}_R = \sigma(R_*)/\sqrt{2R_*^2k(R_*)}$ using \Eqsref{eq:Req} and \eqref{eq:mobility}, where we linearize $v(R)\approx-k(R-R_*)$ about the stationary radius $R_*$.
Simulations solve \Eqsref{eq:continuity}--\eqref{eq:reaction} on a spherically symmetric 3D grid of size $1000w$ and resolution $0.5w$ using \texttt{py-pde}~\cite{Zwicker2020} (Details and parameters in Endmatter).
}\label{fig:phenemenology}
\end{figure}

We consider an incompressible, binary fluid, comprising components A and B, as a minimal model for chemically active droplets.
The system's state is specified by the volume fraction field $\varphi(\vec{r})$ of component A, whereas component B occupies the fraction $1-\varphi(\vec{r})$.
The evolution follows from the continuity equation,
\begin{equation}\label{eq:continuity}
        \frac{\partial\varphi}{\partial t}+\vec{\nabla}\cdot\vec{J}=s
        \;,
\end{equation}
which captures the transport of material due to a conservative flux $\vec{J}$, and the conversion between components A and B due to a non-conservative reaction flux $s$.
To ensure that the noise contained in these fluxes reproduces Boltzmann statistics at equilibrium, we interpret \Eqref{eq:continuity} in the anti-Itô convention~\cite{lanccon2001drift,lau2007state}.

The conservative flux %
\begin{equation}
    \vec{J}=-M\vec{\nabla}\mu+\sqrt{2\kBT\nu M} \,\vec{\eta}
    \label{eqn:diffusive_flux}
\end{equation}
encodes diffusive transport proportional to the density-dependent mobility $M(\varphi)$.
The second term of \Eqref{eqn:diffusive_flux} denotes thermal noise, where $\kBT $ is the thermal energy, $\nu$ is a molecular volume, and $\vec{\eta}$ is a vector of uncorrelated Gaussian white noises with zero mean and unit variance.
In contrast, the deterministic flux is driven by gradients in the exchange chemical potential $\mu=\nu \, \delta F[\varphi]/\delta\varphi$.
This flux redistributes material to minimize the free energy
\begin{equation}\label{eq:F}
    F[\varphi]=\int\left[f(\varphi)+\frac{\kappa}{2}\lvert\nabla\varphi\rvert^2\right]\diff V\; ,
\end{equation}
where $f(\varphi)$ denotes the free energy density and $\kappa$ penalizes composition gradients.

In the absence of chemical reactions ($s=0$), \Eqref{eq:continuity} conserves material and a suitable choice of $f(\varphi)$ can lead to a phase-separated equilibrium state where dense and dilute phases with respective volume fraction $\varphi_\mathrm{in}^{(0)}$  and $\varphi_\mathrm{out}^{(0)}$ coexist in the thermodynamic limit.
In finite systems, the dense phase can form a spherical droplet of radius $R$ [Fig.~\ref{fig:phenemenology}(a)], which is separated from the dilute surroundings by an interface with surface tension $\gamma\sim\sqrt{\kappa b}$ and width $w\sim\sqrt{\kappa/b}$, where $b$ is a characteristic energy density scale of $f(\varphi)$.
The corresponding coexisting compositions are then slightly elevated  due to Laplace pressure, $\varphi^\mathrm{eq}_\mathrm{in/out}\approx\varphi^{(0)}_\mathrm{in/out}+2\gamma/[(\varphi^{(0)}_\mathrm{in}-\varphi^{(0)}_\mathrm{out})f''(\varphi^{(0)}_\mathrm{in/out})R]$~\cite{thomson18724,weber2019physics,zwickerPhysicsDropletRegulation2025}.
Without chemical reactions, the long-term dynamics minimize $F$ and only a single droplet, whose size scales with system size, exists.

To describe the chemical interconversion between components A and B, we use linear non-equilibrium thermodynamics to express the non-conservative reaction flux,
\begin{equation}\label{eq:reaction}
    s=-\Lambda_\mathrm{p}\mu-\Lambda_\mathrm{a}(\mu+\Delta\mu)+\sqrt{2\kBT\nu (\Lambda_\mathrm{p}+\Lambda_\mathrm{a})}\,\xi\;,
\end{equation}
which is driven by chemical potential difference $\mu$ between A and B. 
We distinguish a \textit{passive} reaction with mobility $\Lambda_\mathrm{p}(\varphi)$ from an \textit{active} reaction with mobility $\Lambda_\mathrm{a}(\varphi)$, which requires an energy input $\Delta\mu$ that we assume to be continually  maintained.
The reaction mobilities can depend on composition, for example through the presence of enzymes that might accumulate inside or outside the droplet~\cite{kirschbaum2021controlling}.
Importantly, the \textit{active} reaction maintains the system away from equilibrium, so $F$ is no longer minimized.
Instead, active reactions can enable size-controlled droplets that arrest or even reverse Ostwald ripening \cite{zwicker2015suppression,wurtz2018chemical,kirschbaum2021controlling,weber2019physics,li2020non,zwickerPhysicsDropletRegulation2025,kusters2025size}.
Finally, both reaction pathways contribute to the total reaction noise described by the last term in \Eqref{eq:reaction}, where $\xi$ is another uncorrelated Gaussian white noise with zero mean and unit variance. %
Taken together, \Eqsref{eq:continuity}--\eqref{eq:reaction} fully describe our model for the dynamics of fluctuating chemically active droplets.

To focus on size fluctuations as the lowest-order perturbation, we enforce spherical symmetry and simulate the concentration field $\varphi(r, t)$ to determine the stochastic trajectory $R(t)$ of the droplet radius [Fig.~\ref{fig:phenemenology}(b)].
Fig.~\ref{fig:phenemenology}(c) shows that the coefficient of variation in the stationary state strongly depends on model parameters, such as the active reaction mobility $\Lambda_\mathrm{a}$.
Evidently, stronger active reactions lead to much larger variations in droplet size, suggesting that activity amplifies fluctuations.

To understand how activity affects  size fluctuations of chemically active droplets, we aim to derive a Langevin equation in anti-Itô convention for the droplet radius~$R$,
\begin{equation}\label{eq:stoch}
    \frac{\diff R}{\diff t} = v(R) + \sigma(R) \zeta \; ,
\end{equation}
where $v(R)$ denotes the deterministic drift velocity, $\sigma(R)$ denotes the noise amplitude, and $\zeta$ is an uncorrelated Gaussian white noise with zero mean and unit variance.
To achieve this, we first briefly review the deterministic dynamics~\cite{weber2019physics,zwickerPhysicsDropletRegulation2025,kusters2025size}, and then extend our analysis to fluctuating droplets.

\mysubsection{Deterministic growth}\label{sec:thin-interface}%
To study the deterministic growth of spherical droplets, we investigate a large, isolated droplet obeying $R\gg w$.
Moreover, since reactions are weak, they do not affect interfacial properties like the surface tension~$\gamma$, the interface width~$w$, and the coexisting compositions $\varphi_\mathrm{in}^\mathrm{eq}$ and $\varphi_\mathrm{out}^\mathrm{eq}$. %
We can then derive an effective thin-interface model, where chemical reactions only contribute in the bulk phases, and the induced diffusive fluxes determine the droplet growth velocity~\cite{bray1994theory},
\begin{equation}\label{eq:R1}
    v\approx
    \frac{1}{\Delta\varphi}\left(-\left[M\partial_r\mu\right]_{r\uparrow R}+\left[M\partial_r\mu\right]_{r\downarrow R}\right)\;, 
\end{equation}
where $\Delta\varphi\approx\varphi_\mathrm{in}^{(0)}-\varphi_\mathrm{out}^{(0)}$. 
To obtain the chemical potential profile $\mu(r)$, we determine $\varphi(r)$ inside ($\alpha=\mathrm{in}$) and outside ($\alpha=\mathrm{out}$) the droplet by recasting \Eqref{eq:continuity} as
\begin{equation}\label{eq:piecewise_Eqn}
      \frac{\partial \varphi_\alpha}{\partial t} \approx M_\alpha\left(\nabla^2\mu_\alpha-\frac{1}{\lp{\alpha}^2}\mu_\alpha-\frac{1}{\la{\alpha}^2}(\mu_\alpha+\Delta\mu)\right)
      \;,
\end{equation}
where we have assumed that diffusive and reaction mobilities do not vary much within either bulk phase.
Consequently, we have used $M_\alpha=M(\varphi_\alpha^{(0)})$, and identify the passive and active reaction-diffusion lengths $l_{p,\alpha}=[M_\alpha/\Lambda_\mathrm{p}(\varphi_\alpha^{(0)})]^{1/2}$ and $l_{a,\alpha}=[M_\alpha/\Lambda_\mathrm{a}(\varphi_\alpha^{(0)})]^{1/2}$, respectively. %
Since the system is large, we impose Neumann conditions at infinity. 
At the interface at $r=R$, we impose equilibrium phase coexistence: $\varphi_\mathrm{in}=\varphi_\mathrm{in}^\mathrm{eq}$, $\varphi_\mathrm{out}=\varphi_\mathrm{out}^\mathrm{eq}$, and $\mu_\mathrm{in}=\mu_\mathrm{out}=\nu(\Delta f + 2\gamma R^{-1})/\Delta\varphi$, where 
$\Delta f \approx f(\varphi^{(0)}_\mathrm{in}) - f(\varphi^{(0)}_\mathrm{out})$ approximates the difference in free energy density between the phases~\cite{thomson18724,bray1994theory,weber2019physics,zwickerPhysicsDropletRegulation2025}.
Taken together, \Eqsref{eq:R1}--\eqref{eq:piecewise_Eqn} fully specify the deterministic droplet kinetics.

To derive an explicit expression for the velocity~$v(R)$ in \Eqref{eq:stoch}, we assume as usual that the droplet interface moves slowly compared to the relaxation of the volume fraction field~$\varphi$ \cite{bray1994theory}.
We can then insert the steady solution of \Eqref{eq:piecewise_Eqn} into \Eqref{eq:R1}, to obtain the quasi-static solution~\cite{kusters2025size}
\begin{multline}
    v(R)\approx-\frac{1}{R\Delta\varphi^2}\biggl[M_\mathrm{out}\left(\varepsilon_\mathrm{out}+\frac{2\gamma\nu}{R}\right)\left(1+\frac{R}{l_\mathrm{out}}\right)
\\
    +M_\mathrm{in}\left(\varepsilon_\mathrm{in}+\frac{2\gamma\nu}{R}\right)\left(\frac{R}{l_\mathrm{in}}\coth\!\left({\frac{R}{l_\mathrm{in}}}\right)-1\right)\biggr]
    \;,
\label{eq:Req}
\end{multline}
where we have defined effective reaction-diffusion lengths $l_\alpha= [\la{\alpha}^{-2}+\lp{\alpha}^{-2}]^{-\frac12}$, which quantify the typical distance a component diffuses in phase~$\alpha$ before undergoing any reaction.
We have also introduced energy scales $\varepsilon_\alpha=\Delta\mu\Delta\varphi/(1+\la{\alpha}^2\lp{\alpha}^{-2})+\nu \Delta f$ representing the effective energy that drives reactions in phase~$\alpha$.
Accordingly, the first term in \Eqref{eq:Req} stems from reactions that occur in the dilute surroundings of the droplet, whereas the second term stems from reactions inside the dense droplet.
We showed previously that this analytical approximation of the deterministic droplet growth rate is in good agreement with numerical simulations of \Eqsref{eq:continuity}--\eqref{eq:reaction} \cite{kusters2025size}.

\mysubsection{Fluctuating chemically active droplets}\label{sec:active-droplets}

To extend our analysis to fluctuating droplets, we first analyze the \textit{passive limit} ($\Delta\mu=0$).
In the quasistatic limit, the droplet size dynamics are purely relaxational,
\begin{equation}\label{eq:relaxation}
    \frac{\diff R_\mathrm{pass}}{\diff t}=
        -m(R)\frac{\partial U}{\partial R}
        +\sqrt{2\kBT m(R)} \zeta\;,
\end{equation}
where
the system minimizes the relevant potential
\begin{equation}\label{eq:U}
    U(R)\approx \frac{4\pi}{3}R^3 \Delta f
        +4\pi R^2\gamma \;,
\end{equation}
which can be interpreted as a grand potential, since passive chemical reactions effectively model exchange with a particle bath maintained at $\mu=0$.
The second term in \Eqref{eq:U} describes interfacial contributions, which hinder phase separation.
In contrast, the first term captures the thermodynamic drive for phase separation, which we approximate by $\Delta f$.
Finally, the effective mobility $m(R)$ in \Eqref{eq:relaxation} governs the deterministic dynamics and the magnitude of thermal noise, according to the fluctuation-dissipation theorem.
\Eqsref{eq:relaxation} and \eqref{eq:U} thus describe  size fluctuations of a droplet undergoing passive reactions. %

To determine the magnitude of droplet size fluctuations, we next compute the effective mobility $m(R)$ by
comparing \Eqsref{eq:Req}--\eqref{eq:U} in the limit $\Delta\mu=0$.
The result is a noise magnitude $\sigma(R)=\sqrt{2\kBT m(R)}$, given by
\begin{multline}
    \sigma^2(R)\approx \frac{k_\mathrm{B} T \nu}{2\pi R^3\left(\Delta\varphi\right)^2}\Biggl[M_\mathrm{out}\left(1+\frac{R}{l_\mathrm{out}}\right)
\\
    +M_\mathrm{in}\left(\frac{R}{l_\mathrm{in}}\coth\left({\frac{R}{l_\mathrm{in}}}\right)-1\right)\biggr]
    \;,
\label{eq:mobility}
\end{multline}
where we have discarded higher-order corrections due to the $R$-dependence of the equilibrium compositions, which are negligible for strong phase separation.
In the limit of vanishing reactions ($l_\mathrm{in},l_\mathrm{out}\rightarrow\infty$), \Eqref{eq:mobility} reduces to the well-known $R^{-3}$ scaling for model B~\cite{cates2023classical}, although the full dynamics differ because our density is not formally conserved.
We also recover the classical $R^{-2}$ scaling of model A~\cite{chatzittofi2026nonequilibrium} in the limit of small diffusive mobility ($l_\alpha \ll R$), essentially because reactions are limited to the interface~\cite{kusters2025size}.
(However, we do not not recover the full dynamics of model~A because we assumed  only diffusive fluxes move the interface in \Eqref{eq:R1}.)
In any case, \Eqref{eq:mobility} shows that faster chemical reactions (smaller $l_\mathrm{in}$ and $l_\mathrm{out}$) amplify droplet size fluctuations in the passive limit.

We next extend our analysis to fluctuating \textit{active} droplets by assuming that the \textit{passive} noise of \Eqref{eq:mobility} provides a good estimate of the \textit{real} noise spectrum if droplets are weakly driven.
To test this hypothesis, we perform numerical simulations of \Eqsref{eq:continuity}--\eqref{eq:reaction} in the anti-Itô convention.
\figref{fig:phenemenology}(b) shows that the resulting droplet size fluctuations are qualitatively consistent with the Langevin equation~\eqref{eq:stoch} for the droplet radius~$R$, using \Eqsref{eq:Req} and \eqref{eq:mobility} for the deterministic and stochastic contributions, respectively.
Quantitatively, we measure the size fluctuations around the stationary state $R=R_*$, which satisfies $v(R_*)=0$ and $v'(R_*)<0$ [\Eqref{eq:Req}].
Here, the active chemical reactions exert a restoring force that opposes perturbations, which we  approximate as harmonic close to the stationary state: $v(R)\approx-k(R-R_*)$, where $k$ follows from linearizing \Eqref{eq:Req}.
Under these conditions, the size of a chemically active droplet fluctuates around its stationary value with coefficient of variation $\mathrm{CV}_R=\sigma(R_*)/\sqrt{2R_*^2 k(R_*)}$.
To verify our theory, we compare the predicted $\mathrm{CV}_R$ to measurements from simulations of \Eqsref{eq:continuity}--\eqref{eq:reaction} for various parameters (Endmatter), including activities $\Delta\mu$ up to the relevant  energy scale $b\nu$.
The excellent agreement  over many orders of magnitude in $\mathrm{CV}_R$ [Fig.~\ref{fig:phenemenology}(d)] indicates that \Eqsref{eq:stoch}, \eqref{eq:Req}, and \eqref{eq:mobility} describe the size fluctuations of chemically active droplets faithfully.

\begin{figure}
\centering
            \includegraphics[width=\columnwidth]{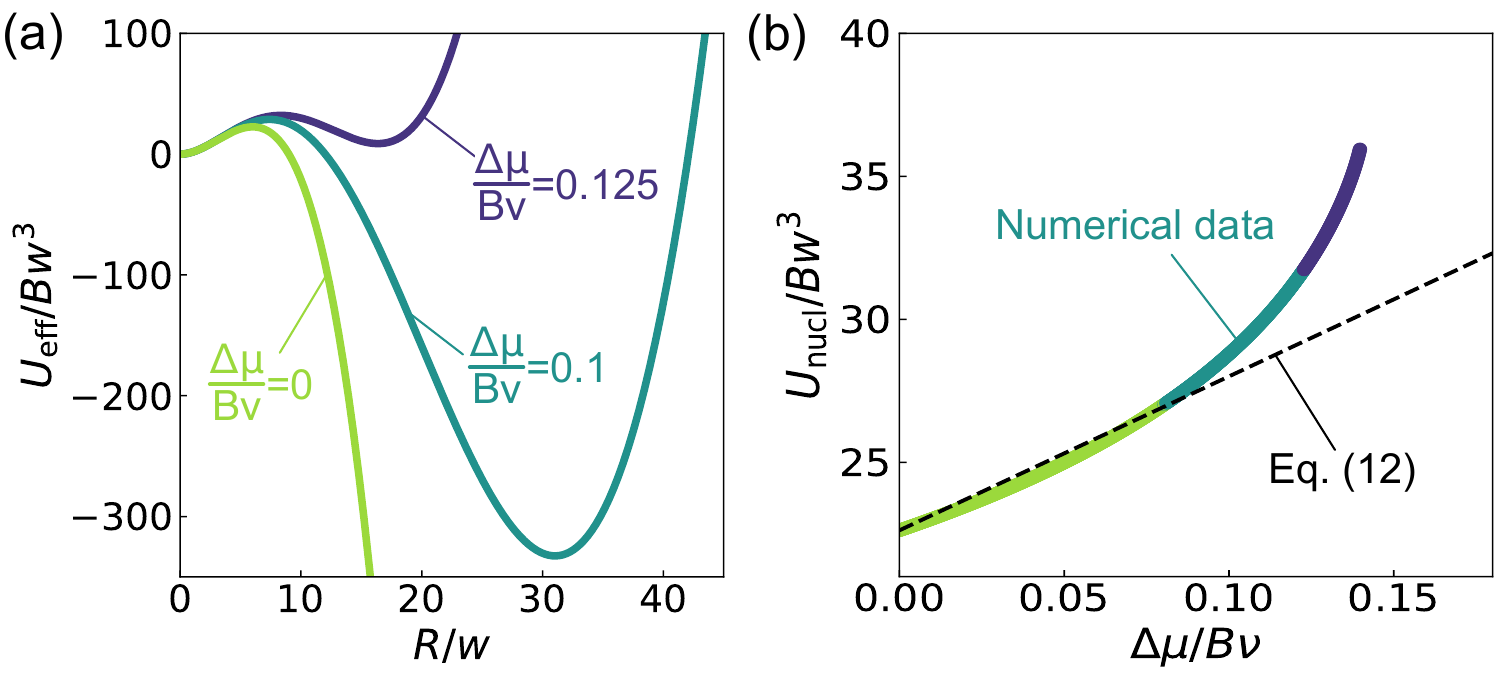}
	\caption{\textbf{Active reactions suppress nucleation.}
    (a) Effective non-equilibrium potential $U_\mathrm{eff}$ as a function of droplet radius $R$ for various  drives~$\Delta\mu$.
    $U_\mathrm{eff}$ either has a single maximum (green line), a maximum and a stable minimum (teal line), or a maximum and a metastable minimum (purple line).
    (b) Nucleation barrier~$U_\mathrm{nucl}$ as a function of $\Delta\mu$ compared to \Eqref{eq:nuc_barrier} (black dashed line).
    Model parameters are $\varepsilon_\mathrm{out}=-0.05\,b\nu$, $\varepsilon_\mathrm{in}=-0.05\,b\nu+\Delta\mu\Delta\varphi$,
    $\gamma=0.15 \, b w$, $l_\mathrm{in}=11 \,w$, $l_\mathrm{out}=18\,w$, and $\Delta\varphi=1$.
    }
	\label{fig:nucleation}
\end{figure}

We next use our stochastic model to derive an effective potential to address droplet nucleation.
Concretely, we hypothesize that we can recast the deterministic velocity~$v(R)$ in terms of relaxation dynamics using the previously derived mobility $m(R)$, $v(R)\approx-m(R)\frac{\partial U_\mathrm{eff}}{\partial R}$.
Solving this equation and integrating defines the effective non-equilibrium potential $U_\mathrm{eff}$ (Endmatter).
In the passive limit [$\Delta\mu=0$, green curve in Fig.~\ref{fig:nucleation}(a)], $U_\mathrm{eff}(R)$ reduces to $U(R)$ given by \Eqref{eq:U}.
For $\Delta f < 0$, $U(R)$ exhibits a maximum defining the nucleation barrier $U_\mathrm{nucl} = \frac{16 \pi}{3}  \gamma ^3/ \Delta f^2$, which determines the nucleation rate $k_\mathrm{nucl}\sim\exp(-U_\mathrm{nucl}/\kBT)$ according to classical nucleation theory~\cite{kalikmanov2013nucleation}.
As expected, nucleation is faster for lower $\gamma$ and larger thermodynamic drive $|\Delta f|$.

Chemical activity ($\Delta\mu\neq0$) generally affects the effective potential $U_\mathrm{eff}$.
As an example, we consider an active reaction that degrades droplet material inside the droplet in Fig.~\ref{fig:nucleation}(a).
Depending on the drive $\Delta\mu$, $U_\mathrm{eff}(R)$ either exhibits a single maximum like the equilibrium potential (green curve), a maximum and a stable minimum that corresponds to the steady state droplet radius~$R_*$ (teal curve), or a maximum and a metastable minimum (purple curve).
Importantly, the nucleation barrier~$U_\mathrm{nucl}$, determined by the local maximum of $U_\mathrm{eff}(R)$, increases with $\Delta\mu$ [\figref{fig:nucleation}(b)], suggesting that active degradation reactions suppress nucleation.
For weak chemical reactions and low drive $\Delta\mu$, %
\begin{equation}\label{eq:nuc_barrier}
    U_\mathrm{nucl}\approx\frac{16\pi\gamma^3}{3\Delta f^2}\left(1 + \frac{8\pi\gamma^2}{5\nu|\Delta f|^3}\frac{\Delta\mu\Delta\varphi}{l_\mathrm{in}^2}\right),
\end{equation}
consistent with our numerical data [dashed line in \figref{fig:nucleation}(b)].
\Eqref{eq:nuc_barrier} shows that activity ($\Delta\mu>0$) always increases $U_\mathrm{nucl}$ relative to the passive limit.
This increase is proportional to the internal reaction rate~$l_\mathrm{in}^{-2} \propto \Lambda_\mathrm{a}(\varphi^{(0)}_\mathrm{in})+ \Lambda_\mathrm{p}(\varphi^{(0)}_\mathrm{in})$, consistent with chemically active droplets with linearized reactions~\cite{ziethen2023nucleation}.
Essentially, this suppression of nucleation is caused by active degradation of droplet material in dense regions, which opposes any local increases in concentration.

\mysection{Conclusion and discussion}\label{sec:conclusion}

In summary, we have shown that fluctuating chemically active droplets can be accurately described by a thin-interface model in the regime of weak reactions and strong phase separation.
This description accurately predicts deterministic droplet growth, and a stochastic noise term derived from the equivalent passive system quantitatively captures droplet size fluctuations at weak or moderate activity.
We also harness our theory to infer the effective non-equilibrium potential that governs droplet dynamics.
For the concrete case where active reactions degrade material inside the droplet, we find amplified size fluctuations [\Eqref{eq:mobility}], but also elevated nucleation barriers [\Eqref{eq:nuc_barrier}], revealing opposite effects on the nucleation rate.
However, numerical simulations~\cite{ziethen2023nucleation} suggest that such reactions generally suppress nucleation.

Our passive-noise approximation implicitly assumed that the interfacial profile retains its quasistatic shape, so the agreement of the noise spectrum with numerical simulations of the full system validates this assumption at the same level.
This contrasts with purely nonconserved active systems, where shape deviations modify the mobility \cite {chatzittofi2026nonequilibrium, ziethen2026nucleationtimereversalsymmetrybreaking}.

Our work provides a quantitative description of droplet size fluctuations in a regime that is biologically and technologically relevant.
Pragmatically, our theory also enables significantly more efficient simulations that include noise at the droplet level \cite{kulkarni2023effective}. %
In addition, we provide a foundation for more elaborate stochastic studies into shape fluctuations and droplet-droplet correlations, as well as for investigations into the large-activity and large-noise regimes.

\textit{Acknowledgement}---%
We thank Filipe Thewes for helpful discussions.
We gratefully acknowledge support and funding from the Max Planck Society, the European Union (ERC,
EmulSim, No.~101044662), and the EPSRC through grant EP/Z534766/1.

\textit{Data availability}---%
The data that support the findings of this study were generated by numerical simulations.
The source code of the simulations will be made publicly available on Zenodo.

\bibliography{Bibliography.bib}

\appendix

\section{Details of numerical simulations}
\label{sec:numerical_details}

We perform numerical simulations of \Eqsref{eq:continuity}--\eqref{eq:reaction} in a spherically symmetric 3D system of radial size $1000w$ and resolution $0.5w$ using \texttt{py-pde}~\cite{Zwicker2020}.
For concreteness, we use constant diffusive mobility $M$ and the Ginzburg-Landau free-energy $f\left(\varphi\right)=\frac{b}{2}\varphi^2(1-\varphi)^2+C_\mathrm{slope}\,b\varphi$ with characteristic energy density scale $b$ and internal energy difference $C_\mathrm{slope}$ between components A and B.
We parametrize the composition-dependent reaction rates as $\Lambda_\mathrm{p}(\varphi)=\lambda_\mathrm{p}[1-\tanh(\frac{\varphi-0.5}{0.1})]Mb\nu w^{-2}$ and $\Lambda_\mathrm{a}(\varphi)=\lambda_\mathrm{a}[1+\tanh(\frac{\varphi-0.5}{0.1})]Mb\nu w^{-2}$, where $\nu$ is a molecular volume scale and $w$ the interface width.
This choice guarantees a smooth interpolation between the inside and outside of the droplet, where both reaction rates are approximately constant.
In addition, we parametrize the active drive as $\Delta\mu=C_\mathrm{drive}\,b\nu$ and the thermal energy $k_\mathrm{B}T= C_{\mathrm{noise}}\,bw^3$, where $C_\mathrm{drive}$ and $C_{\mathrm{noise}}$ are adjustable model parameters.
For \figref{fig:phenemenology}(b,c), we set $C_\mathrm{slope}=-0.05$, $C_\mathrm{drive}=0.1$, $\lambda_\mathrm{p}=0.0015$, $\lambda_\mathrm{a}=0.0042$, $C_\mathrm{noise}=0.5$,  and $\tau=\kappa/Mb^2$.
For \figref{fig:phenemenology}(d), we chose many combinations of parameters to sample the system widely. 
Concretely, we chose $C_\mathrm{slope}\in\{-0.01,\,-0.05,\,-0.09\}$, $C_\mathrm{drive}\in\{0.1,\,0.18,\,1\}$, $5\times10^{-6}<\lambda_\mathrm{p}<10^{-1}$, $8\times10^{-6}<\lambda_\mathrm{a}<4\times10^{-2}$, and $C_\mathrm{noise}\in\{0.001,\,0.0315,\,1.0\}$.
Here, the range of passive and active reaction rates is chosen by for each combination of $C_\mathrm{slope}$ and $C_\mathrm{drive}$
generating $50\times50$ combinations of $\lambda_\mathrm{p}^{-1}\in\{0,\frac{C_\mathrm{drive}\Delta\varphi+C_\mathrm{slope}}{2\sigma}\}$ and $\lambda_\mathrm{a}^{-1}\in\{0,-\frac{C_\mathrm{slope}}{2\sigma}\}$, sampled equidistantly.
In this manner, we include stationary states up to the critical reaction rates beyond which size control breaks down~\cite{kusters2025size}.
Since the noise amplitude does not significantly affect the stationary state, we use the same combinations of the reaction rates $\lambda_p^{-1}$ and $\lambda_a^{-1}$ for all values of $C_\mathrm{noise}$.
We then retain only those parameter combinations that have a stationary-state droplet radius~$R_*$ within the range $0<R_*<100\,w$.

To measure stationary-state size fluctuations from simulations, we first perform a deterministic simulation to reach the stationary state.
We subsequently use this state as the initialization for a stochastic simulation of length $t_\mathrm{sim}=2\times10^{6} \, \tau$ and time step $\mathrm{d}t_\mathrm{sim}=4\times10^{-4} \, \tau$.
At each time step, we determine the droplet radius~$R$ implicitly via $\varphi(R)=\frac12$.
We then measure the time-averaged size fluctuations by computing the squared variations in droplet radius, $\langle [R(t+\Delta t)-R(t)]^2\rangle$, which converges to twice the droplet radius variance, $2(\langle R^2\rangle-\langle R\rangle^2)$, in the limit of large lag time~$\Delta t$.
Note that this time average coincides with the ensemble average of the system for the same model parameters since we measure fluctuations about the stationary state.
Our data points correspond to the average of the plateau value taken over the range of lag times $5\times10^{4}\,\tau\leq\Delta t\leq 10^5\,\tau$, and the error bars correspond to the standard deviation.

In Fig.~\ref{fig:phenemenology}(d), we plot only those data points that have reached the plateau of the autocorrelation function within the simulation time, which we test for by dividing the range $5\times10^{4}\,\tau\leq\Delta t\leq 10^5\,\tau$ into $5$ temporal blocks and evaluating the mean of the autocorrelation function in each.
For a data point to be categorized as plateau-reaching, there must be no systematic trend between the block means (no more than 20\% of subsequent blocks can have increasing or decreasing means), and the relative trend (difference in mean between last and first block divided by the average of block means) and spread (standard deviation of block means divided by average of block means) must be below the threshold $0.1$.

\section{Effective non-equilibrium potential}
\label{sec:Ueff}
We obtain the effective non-equilibrium potential by equating the deterministic term in \Eqref{eq:relaxation}, $-m(R)\frac{\partial U_\mathrm{eff}}{\partial R}$, where $m(R)$ is given by \Eqref{eq:mobility}, to the deterministic drift velocity $v(R)$, as given by \Eqref{eq:Req}.
Solving for $\partial U_\mathrm{eff}/\partial R$ and integrating yields
\begin{widetext}\label{eq:eff_potential}
\begin{multline}
    U_\mathrm{eff}(R)=\frac{2\pi}{3\nu(l_\mathrm{in}-l_\mathrm{out})(l_\mathrm{in}+l_\mathrm{out})}\Bigg\{R\Bigg[(l_\mathrm{in}-l_\mathrm{out})R(-3l_\mathrm{in}l_\mathrm{out}\varepsilon_\mathrm{in}+2l_\mathrm{out}R\varepsilon_\mathrm{in}+3l_\mathrm{in}l_\mathrm{out}\varepsilon_\mathrm{out}+2l_\mathrm{in}R\varepsilon_\mathrm{out}+6(l_\mathrm{in}+l_\mathrm{out}\gamma\nu)\\
    +6l_\mathrm{in}^2l_\mathrm{out}(l_\mathrm{out}+R)(\varepsilon_\mathrm{in}-\varepsilon_\mathrm{out})\ln\left(1+e^{-\frac{2R}{l_\mathrm{in}}}\frac{l_\mathrm{out}-l_\mathrm{in}}{l_\mathrm{in}+l_\mathrm{out}}\right)\Bigg]-3l_\mathrm{in}^3l_\mathrm{out}(\varepsilon_\mathrm{in}-\varepsilon_\mathrm{out})\times\\
    \sum_{n=1}^\infty\Bigg[\frac{l_\mathrm{out}+2R}{n^2}\left(e^{-\frac{2R}{l_\mathrm{in}}}\frac{l_\mathrm{in}-l_\mathrm{out}}{l_\mathrm{in}+l_\mathrm{out}}\right)^n-\frac{l_\mathrm{out}}{n^2}\left(\frac{2l_\mathrm{in}}{l_\mathrm{in}+l_\mathrm{out}}-1\right)^n\\
    +\frac{l_\mathrm{in}}{n^3}\left(\left(e^{-\frac{2R}{l_\mathrm{in}}}\frac{l_\mathrm{in}-l_\mathrm{out}}{l_\mathrm{in}+l_\mathrm{out}}\right)^n-\left(\frac{2l_\mathrm{in}}{l_\mathrm{in}+l_\mathrm{out}}-1\right)^n\right)\Bigg]\Bigg\},
\end{multline}
\end{widetext}
where we have set $M_\mathrm{in}=M_\mathrm{out}$ and chosen the integration constant such that $U_\mathrm{eff}(0)=0$.

To derive a tractable expression for the nucleation barrier, we focus on the case where the external reaction is passive ($\varepsilon_\mathrm{out}=-|\Delta f|\nu$) and the internal reaction is active ($\varepsilon_\mathrm{out}=-|\Delta f|\nu+\Delta\mu\Delta\varphi$). 
We first expand the non-equilibrium potential to second order in $R$ around the passive nucleation radius, $R_\mathrm{nucl}^\mathrm{pass}=2\gamma\nu/|\Delta f|$, and use the resulting quadratic equation in $R$ to solve for the nucleation barrier in the active system.
We then expand the calculated nucleation barrier to first order in $\Delta\mu$ and to second order in $l_\mathrm{in}^{-1}$ and $l_\mathrm{out}^{-1}$ to arrive at \Eqref{eq:nuc_barrier} in the main text.

\end{document}